
\font \eightbf         = cmbx8
\font \eighti          = cmmi8
\font \eightit         = cmti8
\font \eightrm         = cmr8
\font \eightsl         = cmsl8
\font \eightsy         = cmsy8
\font \eighttt         = cmtt8
\font \tenbf           = cmbx9
\font \teni            = cmmi9
\font \tenit           = cmti9
\font \tenrm           = cmr9
\font \tensl           = cmsl9
\font \tensy           = cmsy9
\font \tentt           = cmtt9

\font \kleinhalbcurs   = cmmib10 scaled 800

\font \sixbf           = cmbx6
\font \sixi            = cmmi6
\font \sixrm           = cmr6
\font \sixsy           = cmsy6
\font \tafonts         = cmbx12
\font \tafontss        = cmbx10
\font \tafontt         = cmbx10 scaled\magstep2
\font \tams            = cmmib10
\font \tenmib          = cmmib10
\font \tamt            = cmmib10
\font \tass            = cmsy10
\font \tasss           = cmsy7
\font \tast            = cmsy10 scaled\magstep2
\font \tbfonts         = cmbx8
\font \tbfontss        = cmbx10  scaled 667
\font \tbfontt         = cmbx10 scaled\magstep1
\font \tbmt            = cmmib10
\font \tbss            = cmsy8
\font \tbsss           = cmsy6
\font \tbst            = cmsy10  scaled\magstep1
\vsize=23.5truecm
\hoffset=-1true cm
\voffset=-1true cm
\newdimen\fullhsize
\fullhsize=40cc
\hsize=19.5cc
\def\fullline{\hbox to\fullhsize}
\def\makefootline{\baselineskip=10dd \fullline{\the\footline}}
\def\makeheadline{\vbox to 0pt{\vskip-22.5pt
            \fullline{\vbox to 8.5pt{}\the\headline}\vss}\nointerlineskip}
\let\lr=L \newbox\leftcolumn
\output={\global\topskip=10pt
         \if L\lr
            \global\setbox\leftcolumn=\columnbox \global\let\lr=R
            \message{[left\the\pageno]}%
            \ifnum\pageno=1
               \global\topskip=\fullhead\fi
         \else
            \doubleformat \global\let\lr=L
         \fi
         \ifnum\outputpenalty>-2000 \else\dosupereject\fi}
\def\doubleformat{\shipout\vbox{\makeheadline
    \fullline{\box\leftcolumn\hfil\columnbox}
           \makefootline}
           \advancepageno}
\def\columnbox{\leftline{\pagebody}}
\outer\def\bye{\bigskip\typeset
\sterne=1\ifx\speciali\undefined\else
\loop\smallskip\noindent special character No\number\sterne:
\csname special\romannumeral\sterne\endcsname
\advance\sterne by 1\global\sterne=\sterne
\ifnum\sterne<11\repeat\fi
\if R\lr\null\fi\vfill\supereject\end}
\def\typeset{\begpet\noindent This article was processed by the author using
Sprin\-ger-Ver\-lag \TeX\ AA macro package 1989.\endpet}
\hfuzz=2pt
\vfuzz=2pt
\tolerance=1000
\fontdimen3\tenrm=1.5\fontdimen3\tenrm
\fontdimen7\tenrm=1.5\fontdimen7\tenrm
\abovedisplayskip=3 mm plus6pt minus 4pt
\belowdisplayskip=3 mm plus6pt minus 4pt
\abovedisplayshortskip=0mm plus6pt
\belowdisplayshortskip=2 mm plus4pt minus 4pt
\predisplaypenalty=0
\clubpenalty=20000
\widowpenalty=20000
\parindent=1.5em
\frenchspacing
\def\newline{\hfill\break}%
\nopagenumbers
\def\AALogo{\setbox254=\hbox{ ASTROPHYSICS }%
\vbox{\baselineskip=10dd\hrule\hbox{\vrule\vbox{\kern3pt
\hbox to\wd254{\hfil ASTRONOMY\hfil}
\hbox to\wd254{\hfil AND\hfil}\copy254
\hbox to\wd254{\hfil\number\day.\number\month.\number\year\hfil}
\kern3pt}\vrule}\hrule}}
\def\paglay{\headline={{\tenrm\hsize=.75\fullhsize\ifnum\pageno=1
\vbox{\baselineskip=10dd\hrule\line{\vrule\kern3pt\vbox{\kern3pt
\hbox{\bf A and A Manuskript-Nr.}
\hbox{(will be inserted by hand later)}
\kern3pt\hrule\kern3pt
\hbox{\bf Your thesaurus codes are:}
\hbox{\rightskip=0pt plus3em\advance\hsize by-7pt
\vbox{\noindent\ignorespaces\the\THESAURUS}}
\kern3pt}\hfil\kern3pt\vrule}\hrule}
\rlap{\quad\AALogo}\hfil
\else\ifodd\pageno\hfil\folio\else\folio\hfil\fi\fi}}}
\ifx \undefined\instruct
\headline={\tenrm\ifodd\pageno\hfil\folio
\else\folio\hfil\fi}\fi
\newcount\eqnum\eqnum=0
\def\autnum{\global\advance\eqnum by 1{\rm(\the\eqnum)}}
\newtoks\eq\newtoks\eqn
\catcode`@=11
\def\eqalign#1{\null\vcenter{\openup\jot\m@th
  \ialign{\strut\hfil$\displaystyle{##}$&$\displaystyle{{}##}$\hfil
      \crcr#1\crcr}}}
\def\displaylines#1{{}$\displ@y
\hbox{\vbox{\halign{$\@lign\hfil\displaystyle##\hfil$\crcr
    #1\crcr}}}${}}
\def\eqalignno#1{{}$\displ@y
  \hbox{\vbox{\halign to\hsize{\hfil$\@lign\displaystyle{##}$\tabskip\z@skip
    &$\@lign\displaystyle{{}##}$\hfil\tabskip\centering
    &\llap{$\@lign##$}\tabskip\z@skip\crcr
    #1\crcr}}}${}}
\def\leqalignno#1{{}$\displ@y
\hbox{\vbox{\halign
to\hsize{\qquad\hfil$\@lign\displaystyle{##}$\tabskip\z@skip
    &$\@lign\displaystyle{{}##}$\hfil\tabskip\centering
    &\kern-\hsize\rlap{$\@lign##$}\tabskip\hsize\crcr
    #1\crcr}}}${}}
\def\generaldisplay{%
\ifeqno
       \ifleqno\leftline{$\displaystyle\the\eqn\quad\the\eq$}%
       \else\line{$\displaystyle\the\eq\hfill\the\eqn$}\fi
\else
       \leftline{$\displaystyle\the\eq$}%
\fi
\global\eq={}\global\eqn={}}%
\newif\ifeqno\newif\ifleqno \everydisplay{\displaysetup}
\def\displaysetup#1$${\displaytest#1\eqno\eqno\displaytest}
\def\displaytest#1\eqno#2\eqno#3\displaytest{%
\if!#3!\ldisplaytest#1\leqno\leqno\ldisplaytest
\else\eqnotrue\leqnofalse\eqn={#2}\eq={#1}\fi
\generaldisplay$$}
\def\ldisplaytest#1\leqno#2\leqno#3\ldisplaytest{\eq={#1}%
\if!#3!\eqnofalse\else\eqnotrue\leqnotrue\eqn={#2}\fi}
\catcode`@=12 
\mathchardef\Gamma="0100
\mathchardef\Delta="0101
\mathchardef\Theta="0102
\mathchardef\Lambda="0103
\mathchardef\Xi="0104
\mathchardef\Pi="0105
\mathchardef\Sigma="0106
\mathchardef\Upsilon="0107
\mathchardef\Phi="0108
\mathchardef\Psi="0109
\mathchardef\Omega="010A

\def\utw{\smash{\rlap{\lower5pt\hbox{$\sim$}}}}
\def\udtw{\smash{\rlap{\lower6pt\hbox{$\approx$}}}}

\def\diameter{{\ifmmode\mathchoice
{\ooalign{\hfil\hbox{$\displaystyle/$}\hfil\crcr
{\hbox{$\displaystyle\mathchar"20D$}}}}
{\ooalign{\hfil\hbox{$\textstyle/$}\hfil\crcr
{\hbox{$\textstyle\mathchar"20D$}}}}
{\ooalign{\hfil\hbox{$\scriptstyle/$}\hfil\crcr
{\hbox{$\scriptstyle\mathchar"20D$}}}}
{\ooalign{\hfil\hbox{$\scriptscriptstyle/$}\hfil\crcr
{\hbox{$\scriptscriptstyle\mathchar"20D$}}}}
\else{\ooalign{\hfil/\hfil\crcr\mathhexbox20D}}%
\fi}}

\normallineskip=1dd
\normallineskiplimit=0dd
\normalbaselineskip=10dd
\textfont0=\tenrm
\textfont1=\teni
\textfont2=\tensy
\textfont\itfam=\tenit
\textfont\slfam=\tensl
\textfont\ttfam=\tentt
\textfont\bffam=\tenbf
\normalbaselines\rm
\def\petit{\def\rm{\fam0\eightrm}
\textfont0=\eightrm \scriptfont0=\sixrm \scriptscriptfont0=\fiverm
 \textfont1=\eighti \scriptfont1=\sixi \scriptscriptfont1=\fivei
 \textfont2=\eightsy \scriptfont2=\sixsy \scriptscriptfont2=\fivesy
 \def\it{\fam\itfam\eightit}%
 \textfont\itfam=\eightit
 \def\sl{\fam\slfam\eightsl}%
 \textfont\slfam=\eightsl
 \def\bf{\fam\bffam\eightbf}%
 \textfont\bffam=\eightbf \scriptfont\bffam=\sixbf
 \scriptscriptfont\bffam=\fivebf
 \def\tt{\fam\ttfam\eighttt}%
 \textfont\ttfam=\eighttt
 \let\tams=\kleinhalbcurs
 \let\tenbf=\eightbf
 \let\sevenbf=\sixbf
 \normalbaselineskip=9dd
 \if Y\REFEREE \normalbaselineskip=2\normalbaselineskip
 \normallineskip=2\normallineskip\fi
 \setbox\strutbox=\hbox{\vrule height7pt depth2pt width0pt}%
 \normalbaselines\rm}%
\def\begpet{\vskip6pt\bgroup\petit}
\def\endpet{\vskip6pt\egroup}
\def\rahmen#1{\vbox{\hrule\line{\vrule\vbox to#1true
cm{\vfil}\hfil\vrule}\vfil\hrule}}
\def\begfig#1cm#2\endfig{\par
   \ifvoid\topins\midinsert\bigskip\vbox{\rahmen{#1}#2}\endinsert
   \else\topinsert\vbox{\rahmen{#1}#2}\endinsert
\fi}
\def\begfigwid#1cm#2\endfig{\par
\if N\lr\else
\if R\lr
\shipout\vbox{\makeheadline
\line{\box\leftcolumn}\makefootline}\advancepageno
\fi\let\lr=N
\topskip=10pt
\output={\plainoutput}%
\fi
\topinsert\line{\vbox{\hsize=\fullhsize\rahmen{#1}#2}\hss}\endinsert}
\def\figure#1#2{\bigskip\noindent{\petit{\bf Fig.\ts#1.\
}\ignorespaces #2\smallskip}}
\def\begtab#1cm#2\endtab{\par
   \ifvoid\topins\midinsert\medskip\vbox{#2\rahmen{#1}}\endinsert
   \else\topinsert\vbox{#2\rahmen{#1}}\endinsert
\fi}
\def\begtabemptywid#1cm#2\endtab{\par
\if N\lr\else
\if R\lr
\shipout\vbox{\makeheadline
\line{\box\leftcolumn}\makefootline}\advancepageno
\fi\let\lr=N
\topskip=10pt
\output={\plainoutput}%
\fi
\topinsert\line{\vbox{\hsize=\fullhsize#2\rahmen{#1}}\hss}\endinsert}
\def\begtabfullwid#1\endtab{\par
\if N\lr\else
\if R\lr
\shipout\vbox{\makeheadline
\line{\box\leftcolumn}\makefootline}\advancepageno
\fi\let\lr=N
\output={\plainoutput}%
\fi
\topinsert\line{\vbox{\hsize=\fullhsize\noindent#1}\hss}\endinsert}

\def\begref{\vskip1cm\begingroup\let\INS=N}
\def\ref{\goodbreak\if N\INS\let\INS=Y\vbox{\noindent\tenbf
References\bigskip}\fi\hangindent\parindent
\hangafter=1\noindent\ignorespaces}
\def\endref{\goodbreak\endgroup}
\def\ack#1{\vskip11pt\begingroup\noindent{\it Acknowledgements\/}.
\ignorespaces#1\vskip6pt\endgroup}

%
%
 \def \aTa  { \goodbreak
     \bgroup
     \par
 \textfont0=\tafontt \scriptfont0=\tafonts \scriptscriptfont0=\tafontss
 \textfont1=\tamt \scriptfont1=\tbmt \scriptscriptfont1=\tams
 \textfont2=\tast \scriptfont2=\tass \scriptscriptfont2=\tasss
     \baselineskip=17dd
     \lineskip=17dd
     \rightskip=0pt plus2cm\spaceskip=.3333em \xspaceskip=.5em
     \pretolerance=10000
     \noindent
     \tafontt}
 %
 \def \eTa{\vskip10pt\egroup
     \noindent
     \ignorespaces}

%
%
%
 %
 \def \aTb{\goodbreak
     \bgroup
     \par
 \textfont0=\tbfontt \scriptfont0=\tbfonts \scriptscriptfont0=\tbfontss
 \textfont1=\tbmt \scriptfont1=\tenmib \scriptscriptfont1=\tams
 \textfont2=\tbst \scriptfont2=\tbss \scriptscriptfont2=\tbsss
     \baselineskip=13dd
     \lineskip=13dd
     \rightskip=0pt plus2cm\spaceskip=.3333em \xspaceskip=.5em
     \pretolerance=10000
     \noindent
     \tbfontt}
 %
 \def \eTb{\vskip10pt
     \egroup
     \noindent
     \ignorespaces}
  %
\catcode`\@=11
\expandafter \newcount \csname c@Tl\endcsname
    \csname c@Tl\endcsname=0
\expandafter \newcount \csname c@Tm\endcsname
    \csname c@Tm\endcsname=0
\expandafter \newcount \csname c@Tn\endcsname
    \csname c@Tn\endcsname=0
\expandafter \newcount \csname c@To\endcsname
    \csname c@To\endcsname=0
\expandafter \newcount \csname c@Tp\endcsname
    \csname c@Tp\endcsname=0
\def \resetcount#1    {\global
    \csname c@#1\endcsname=0}
\def\@nameuse#1{\csname #1\endcsname}
\def\arabic#1{\@arabic{\@nameuse{c@#1}}}
\def\@arabic#1{\ifnum #1>0 \number #1\fi}
 %
\expandafter \newcount \csname c@fn\endcsname
    \csname c@fn\endcsname=0
\def \stepc#1    {\global
    \expandafter
    \advance
    \csname c@#1\endcsname by 1}
\catcode`\@=12
%
%
   \catcode`\@= 11
%
%

\skewchar\eighti='177 \skewchar\sixi='177
\skewchar\eightsy='60 \skewchar\sixsy='60
\hyphenchar\eighttt=-1
\def\footnoterule{\kern-3pt\hrule width 2true cm\kern2.6pt}
\newinsert\footins
\def\footnotea#1{\let\@sf\empty 
  \ifhmode\edef\@sf{\spacefactor\the\spacefactor}\/\fi
  {#1}\@sf\vfootnote{#1}}
\def\vfootnote#1{\insert\footins\bgroup
  \textfont0=\tenrm\scriptfont0=\sevenrm\scriptscriptfont0=\fiverm
  \textfont1=\teni\scriptfont1=\seveni\scriptscriptfont1=\fivei
  \textfont2=\tensy\scriptfont2=\sevensy\scriptscriptfont2=\fivesy
  \interlinepenalty\interfootnotelinepenalty
  \splittopskip\ht\strutbox 
  \splitmaxdepth\dp\strutbox \floatingpenalty\@MM
  \leftskip\z@skip \rightskip\z@skip \spaceskip\z@skip \xspaceskip\z@skip
  \textindent{#1}\footstrut\futurelet\next\fo@t}
\def\fo@t{\ifcat\bgroup\noexpand\next \let\next\f@@t
  \else\let\next\f@t\fi \next}
\def\f@@t{\bgroup\aftergroup\@foot\let\next}
\def\f@t#1{#1\@foot}
\def\@foot{\strut\egroup}
\def\footstrut{\vbox to\splittopskip{}}
\skip\footins=\bigskipamount 
\count\footins=1000 
\dimen\footins=8in 
   \def \bfootax  {\bgroup\tenrm
                  \baselineskip=12pt\lineskiplimit=-6pt
                  \hsize=19.5cc
                  \def\textindent##1{\hang\noindent\hbox
                  to\parindent{##1\hss}\ignorespaces}%
                  \footnotea{$^\star$}\bgroup}
   \def \efootax  {\egroup\egroup}
   \def \bfootay  {\bgroup\tenrm
                  \baselineskip=12pt\lineskiplimit=-6pt
                  \hsize=19.5cc
                  \def\textindent##1{\hang\noindent\hbox
                  to\parindent{##1\hss}\ignorespaces}%
                  \footnotea{$^{\star\star}$}\bgroup}
   \def \efootay  {\egroup\egroup }
   \def \bfootaz {\bgroup\tenrm
                  \baselineskip=12pt\lineskiplimit=-6pt
                  \hsize=19.5cc
                  \def\textindent##1{\hang\noindent\hbox
                  to\parindent{##1\hss}\ignorespaces}%
                 \footnotea{$^{\star\star\star}$}\bgroup}
   \def \efootaz {\egroup \egroup}
\def\fonote#1{\mehrsterne$^{\the\sterne}$\begingroup
       \def\textindent##1{\hang\noindent\hbox
       to\parindent{##1\hss}\ignorespaces}%
\vfootnote{$^{\the\sterne}$}{#1}\endgroup}
\catcode`\@=12
%
\everypar={\let\lasttitle=N\everypar={\parindent=1.5em}}%
%
%
\def \titlea#1{\stepc{Tl}
     \resetcount{Tm}
     \vskip22pt
     \setbox0=\vbox{\vskip 22pt\noindent
     \bf
     \rightskip 0pt plus4em
     \pretolerance=20000
     \arabic{Tl}.\

\textfont1=\tams\scriptfont1=\kleinhalbcurs\scriptscriptfont1=\kleinhalbcurs
     \ignorespaces#1
     \vskip11pt}
     \dimen0=\ht0\advance\dimen0 by\dp0\advance\dimen0 by 2\baselineskip
     \advance\dimen0 by\pagetotal
     \ifdim\dimen0>\pagegoal\eject\fi
     \bgroup
     \noindent
     \bf
     \rightskip 0pt plus4em
     \pretolerance=20000
     \arabic{Tl}.\

\textfont1=\tams\scriptfont1=\kleinhalbcurs\scriptscriptfont1=\kleinhalbcurs
     \ignorespaces#1
     \vskip11pt
     \egroup
     \nobreak
     \parindent=0pt
     \everypar={\parindent=1.5em
     \let\lasttitle=N\everypar={\let\lasttitle=N}}%
     \let\lasttitle=A%
     \ignorespaces}
 %
 \def\titleb#1{\stepc{Tm}
     \resetcount{Tn}
     \if N\lasttitle\else\vskip-11pt\vskip-\baselineskip
     \fi
     \vskip 17pt
     \setbox0=\vbox{\vskip 17pt
     \raggedright
     \pretolerance=10000
     \noindent
     \it
     \arabic{Tl}.\arabic{Tm}.\
     \ignorespaces#1
     \vskip8pt}
     \dimen0=\ht0\advance\dimen0 by\dp0\advance\dimen0 by 2\baselineskip
     \advance\dimen0 by\pagetotal
     \ifdim\dimen0>\pagegoal\eject\fi
     \bgroup
     \raggedright
     \pretolerance=10000
     \noindent
     \it
     \arabic{Tl}.\arabic{Tm}.\
     \ignorespaces#1
     \vskip8pt
     \egroup
     \nobreak
     \let\lasttitle=B%
     \parindent=0pt
     \everypar={\parindent=1.5em
     \let\lasttitle=N\everypar={\let\lasttitle=N}}%
     \ignorespaces}
 %
 \def \titlec#1{\stepc{Tn}
     \resetcount{To}
     \if N\lasttitle\else\vskip-3pt\vskip-\baselineskip
     \fi
     \vskip 11pt
     \setbox0=\vbox{\vskip 11pt
     \noindent
     \raggedright
     \pretolerance=10000
     \arabic{Tl}.\arabic{Tm}.\arabic{Tn}.\
     \ignorespaces#1\vskip6pt}
     \dimen0=\ht0\advance\dimen0 by\dp0\advance\dimen0 by 2\baselineskip
     \advance\dimen0 by\pagetotal
     \ifdim\dimen0>\pagegoal\eject\fi
     \bgroup\noindent
     \raggedright
     \pretolerance=10000
     \arabic{Tl}.\arabic{Tm}.\arabic{Tn}.\
     \ignorespaces#1\vskip6pt
     \egroup
     \nobreak
     \let\lasttitle=C%
     \parindent=0pt
     \everypar={\parindent=1.5em
     \let\lasttitle=N\everypar={\let\lasttitle=N}}%
     \ignorespaces}
 %
 \def\titled#1{\stepc{To}
     \resetcount{Tp}
     \if N\lasttitle\else\vskip-3pt\vskip-\baselineskip
     \fi
     \vskip 11pt
     \bgroup
     \it
     \noindent
     \ignorespaces#1\unskip. \egroup\ignorespaces}
\let\REFEREE=N
\newbox\refereebox
\setbox\refereebox=\vbox
to0pt{\vskip0.5cm\fullline{\hrulefill\tentt\lower0.5ex
\hbox{\kern5pt referee's copy\kern5pt}\hrulefill}\vss}%
\def\refereelayout{\let\REFEREE=M\footline={\copy\refereebox}%
\message{|A referee's copy will be produced}\par
\if N\lr\else
\if R\lr
\shipout\vbox{\makeheadline
\line{\box\leftcolumn}\makefootline}\advancepageno
\fi\let\lr=N
\topskip=10pt
\output={\plainoutput}%
\fi
}
\let\ts=\thinspace
\newcount\sterne \sterne=0
\newdimen\fullhead
\newtoks\RECDATE
\newtoks\ACCDATE
\newtoks\MAINTITLE
\newtoks\SUBTITLE
\newtoks\AUTHOR
\newtoks\INSTITUTE
\newtoks\SUMMARY
\newtoks\KEYWORDS
\newtoks\THESAURUS
\newtoks\SENDOFF
\newlinechar=`\| 
\catcode`\@=\active
\let\INS=N%
\def@#1{\if N\INS $^{#1}$\else\if
E\INS\hangindent0.5\parindent\hangafter=1%
\noindent\hbox to0.5\parindent{$^{#1}$\hfil}\let\INS=Y\ignorespaces
\else\par\hangindent0.5\parindent\hangafter=1
\noindent\hbox to0.5\parindent{$^{#1}$\hfil}\ignorespaces\fi\fi}%
\def\mehrsterne{\advance\sterne by1\global\sterne=\sterne}%
\def\FOOTNOTE#1{\mehrsterne\ifcase\sterne
\or\bfootax \ignorespaces #1\efootax
\or\bfootay \ignorespaces #1\efootay
\or\bfootaz \ignorespaces #1\efootaz\else\fi}%
\def\PRESADD#1{\mehrsterne\ifcase\sterne
\or\bfootax Present address: #1\efootax
\or\bfootay Present address: #1\efootay
\or\bfootaz Present address: #1\efootaz\else\fi}%
\def\maketitle{\paglay%
\def\missing{ ????? }
%
\setbox0=\vbox{\parskip=0pt\hsize=\fullhsize\null\vskip2truecm
\let\kka = \tamt
\edef\test{\the\MAINTITLE}%
\ifx\test\missing\MAINTITLE={MAINTITLE should be given}\fi
\aTa\ignorespaces\the\MAINTITLE\eTa
\let\kka = \tbmt
\edef\test{\the\SUBTITLE}%
\ifx\test\missing\else\aTb\ignorespaces\the\SUBTITLE\eTb\fi
\let\kka = \tams
\edef\test{\the\AUTHOR}%
\ifx\test\missing
\AUTHOR={Name(s) and initial(s) of author(s) should be given}\fi
\noindent{\bf\ignorespaces\the\AUTHOR\vskip4pt}
\let\INS=E%
\edef\test{\the\INSTITUTE}%
\ifx\test\missing
\INSTITUTE={Address(es) of author(s) should be given.}\fi
{\noindent\ignorespaces\the\INSTITUTE\vskip10pt}%
\edef\test{\the\RECDATE}%
\ifx\test\missing
\RECDATE={{\petit $[$the date should be inserted later$]$}}\fi
\edef\test{\the\ACCDATE}%
\ifx\test\missing
\ACCDATE={{\petit $[$the date should be inserted later$]$}}\fi
{\noindent Received \ignorespaces\the\RECDATE\unskip; accepted \ignorespaces
\the\ACCDATE\vskip21pt\bf S}}%
\global\fullhead=\ht0\global\advance\fullhead by\dp0
\global\advance\fullhead by10pt\global\sterne=0
{\parskip=0pt\hsize=19.5cc\null\vskip2truecm
\edef\test{\the\SENDOFF}%
\ifx\test\missing\else\insert\footins{\smallskip\noindent
{\it Send offprint requests to\/}: \ignorespaces\the\SENDOFF}\fi
\hsize=\fullhsize
\let\kka = \tamt
\edef\test{\the\MAINTITLE}%
\ifx\test\missing\message{|Your MAINTITLE is missing.}%
\MAINTITLE={MAINTITLE should be given}\fi
\aTa\ignorespaces\the\MAINTITLE\eTa
\let\kka = \tbmt
\edef\test{\the\SUBTITLE}%
\ifx\test\missing\message{|The SUBTITLE is optional.}%
\else\aTb\ignorespaces\the\SUBTITLE\eTb\fi
\let\kka = \tams
\edef\test{\the\AUTHOR}%
\ifx\test\missing\message{|Name(s) and initial(s) of author(s) missing.}%
\AUTHOR={Name(s) and initial(s) of author(s) should be given}\fi
\noindent{\bf\ignorespaces\the\AUTHOR\vskip4pt}
\let\INS=E%
\edef\test{\the\INSTITUTE}%
\ifx\test\missing\message{|Address(es) of author(s) missing.}%
\INSTITUTE={Address(es) of author(s) should be given.}\fi
{\noindent\ignorespaces\the\INSTITUTE\vskip10pt}%
\edef\test{\the\RECDATE}%
\ifx\test\missing\message{|The date of receipt should be inserted
later.}%
\RECDATE={{\petit $[$the date should be inserted later$]$}}\fi
\edef\test{\the\ACCDATE}%
\ifx\test\missing\message{|The date of acceptance should be inserted
later.}%
\ACCDATE={{\petit $[$the date should be inserted later$]$}}\fi
{\noindent Received \ignorespaces\the\RECDATE\unskip; accepted \ignorespaces
\the\ACCDATE\vskip21pt}}%
\edef\test{\the\THESAURUS}%
\ifx\test\missing\THESAURUS={missing; you have not inserted them}%
\message{|Thesaurus codes are not given.}\fi
\if M\REFEREE\let\REFEREE=Y
\normalbaselineskip=2\normalbaselineskip
\normallineskip=2\normallineskip\normalbaselines\fi
\edef\test{\the\SUMMARY}%
\ifx\test\missing\message{|Summary is missing.}%
\SUMMARY={Not yet given.}\fi
\noindent{\bf Summary. }\ignorespaces
\the\SUMMARY\vskip0.5true cm
\edef\test{\the\KEYWORDS}%
\ifx\test\missing\message{|Missing keywords.}%
\KEYWORDS={Not yet given.}\fi
\noindent{\bf Key words: }\the\KEYWORDS
\vskip3pt\line{\hrulefill}\vfill
\global\sterne=0
\catcode`\@=12}

\def\simle{\lower 2pt \hbox {$\buildrel < \over {\scriptstyle \sim }$}}
\def\simge{\lower 2pt \hbox {$\buildrel > \over {\scriptstyle \sim }$}}

\MAINTITLE={Cosmic rays}

\SUBTITLE{V. The nonthermal radio emission of the old nova GK Per - a signature
of hadronic interactions?}

\AUTHOR={ Peter L.\ts Biermann@1, Richard G. Strom@2, Heino \ts Falcke@1}

\SENDOFF={ P.L.\ts Biermann}


\INSTITUTE={@1 Max Planck Institut f\"ur Radioastronomie, Auf dem H\"ugel
69, D-53121 Bonn, Germany
@2 Netherlands Foundation for Research in Astronomy, Radiosterrenwacht,
P.O.Box 2, NL-7990 AA Dwingeloo, The Netherlands}

\SUMMARY={
Using the model of particle acceleration in shocks in stellar winds developed
by Biermann (1993) and Biermann and Cassinelli (1993) we demonstrate that
the nonthermal radio emission spectrum from the cataclysmic variable binary
star GK Per can be interpreted consistently.  We propose that the radio
spectrum results from hadronic secondary electrons/positrons, which, in
the  optically thin case, give a characteristic radio spectrum with a spectral
index of +1/3 below a maximum and, for a strong shock in a wind slow compared
to the shock speed, of -0.7 above the maximum.  We test the proposed
concept for consistency with the radio data for OB and Wolf Rayet stars as
well as radiosupernovae.  There is some evidence that this process is
important for the radio emission from active galactic nuclei, both in compact
and extended emission.  We argue that such a radio spectrum  is a unique
characteristic for hadronic interactions.}

\KEYWORDS={Acceleration of particles, Accretion disks, Elementary particles,
Close binary stars, Cosmic Rays, Radio continuum: stars}

\THESAURUS={02.01.1, 02.01.2, 02.05.1, 08.02.1, 09.03.2, 13.18.5}

\RECDATE={    }

\ACCDATE={    }

\maketitle

\titlea {Introduction}

\titleb {The origin of cosmic rays}

The origin of cosmic rays as observed near earth is still not fully settled.
Recent work proposes that this origin can be attributed to three major sites,
a) stellar explosions into the interstellar medium (ISM-supernovae),
b) stellar explosions into a former stellar wind (wind-supernovae) , and
c) powerful radio galaxy hot spots.  The predictions were made in Biermann
(1993, paper CR I), and in Rachen and Biermann (1993, paper UHE CR I).  Tests
on a) stellar winds were made in Biermann and Cassinelli (1993, paper
CR II), on b) explosions into a homogeneous interstellar medium in
Biermann and Strom (1993, paper CR III), on c) the chemical abundance and
spectrum against air shower data in Stanev et al. (1993, paper CR IV), and
on d) the ratio of wind-supernovae to ISM-supernovae and the implications
for the chemical abundances of Helium and hydrogen in cosmic rays in Biermann
et al. (1995).  The chemical abundances and  the predicted spectrum for the
extragalactic component (paper UHE CR I)  were checked against air shower
data in Rachen et al. (1993, paper UHE CR II).  This latter check provided
an independent control on the spectrum and chemical composition of the
galactic component beyond the knee, near $5 \, 10^6$ GeV.  A definitive check
on the chemical composition across the knee will probably be possible with the
MACRO-EASTOP experiment  (Battistoni and Navarra, priv. comm.), combining muon
data with  air shower data.  Clearly, any concept of how particles, nuclei and
electrons get injected and accelerated has to account for the nonthermal radio
emission in other sites which are believed  to be basically similar.

\titleb {The nova GK Per}

The nova GK Per, which exploded in 1901 and has been observed ever since with
great care, is such a case.  The radio spectrum of its shell has been
observed  (Reynolds and Chevalier 1984) to be a power law of spectral index
$\approx -2/3$,  which is close to the limiting spectrum of synchrotron
emission  from a shocked region  in a stellar wind in  the case that the
shock is much faster than the wind speed (exactly $-2/3$).  We note that this
radio spectrum is quite unusual for novae  (see, e.g., Hjellming 1990),
since in other cases where radio emission has  been detected, it can readily
be understood as free-free emission.  This is similar to the case of OB and
WR stars (CR II), where free-free radio emission is also very common, and
non-thermal radio emission is clearly less frequent (Abbott et al. 1986,
Bieging et al. 1989).  For OB and WR stars we have suggested that shocks  in
winds accelerate particles to sufficient energies to give the observed
non-thermal radio emission.  In Nath \& Biermann (1994) we have been able to
check the concept of shocks running through the winds of OB stars and then
hitting a surrounding shell with the observed $\gamma$-ray line and continuum
data.

In this paper we try to understand  the radio emission from the shell of the
nova GK Per in order to check the concepts described in the series of papers on
the origin of cosmic rays, and develop them further.

The model which we have developed for strong shocks in stellar winds (see,
MacFarlane \& Cassinelli 1989, CR I and CR II) has been described in some
detail in Biermann (1995).  The model concentrates on spherical
shockwaves, with a magnetic field orientation mostly parallel to the
shock front, as is the case for a spherical shock advancing into an
asymptotic Parker type wind.  We note that in this case a north-south
asymmetry can develop due to the possibility that the sign of the particle
drifts differs in the two hemispheres (towards the pole in one hemisphere and
towards the equator in the other, for instance).  We propose this effect to
explain the asymmetry in the radio emission in the radio shell of GK Per.

Therefore we explore in this paper the idea that the nonthermal radio emission
from the shell produced  by the nova GK Per is also due to a strong shock
initiated in a wind by the outburst of 1901.  For the wind we
assume a Parker-type spiral magnetic field (see, e.g., Jokipii et al. 1977).

\titlea {The low frequency turnover in the radio spectrum}

\titleb {The data for GK Per}

The spectrum of GK Per has been published by Seaquist et al. (1989).  We
have reanalyzed the WSRT data at 49 and 92 cm, making full use of the
redundancy technique to achieve high dynamic range and thereby reduce the
interfering effects of background sources.  This results in an upward
revision of the flux densities $S_{\nu}$ at both wavelengths, and smaller
errors.  The flux density scale is based upon that of Baars et al. (1977),
using 3C48 as a primary, and 3C295 as a secondary calibrator.  The flux
densities are $8.7 \pm 0.5$, $20.6 \pm 1.6$, $29 \pm 3$, $33 \pm 10$,
$38 \pm 6$, and $< 40$ mJy at the frequencies 4.86, 1.49, 0.608, 0.408,
0.327, and 0.151 GHz, respectively.  The improvement over previous data is
the Westerbork points at 0.608 and 0.327 GHz.  Our spectrum is shown in
Fig. 1.  It is clear that the high frequency spectrum turns over,
gradually, somewhere below 1 GHz, suggesting either absorption or a low
energy cutoff in the electron spectrum.

\begfig 6 cm
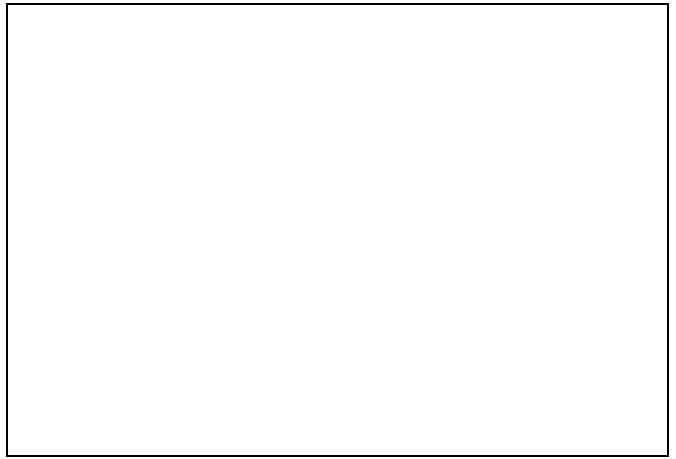
\figure {1} {The newly derived radio spectrum of GK Per using WSRT data in
the redundancy mode.  These newly derived data points are emphasized in this
figure which is otherwise analoguous to the spectrum shown in Seaquist et al.
(1989)}
\endfig

When we fit a powerlaw through the three high frequency data points, we obtain

$$S_{\nu} \; = \; (13.4 \pm 1.3 \, {\rm mJy}) \, ({\nu \over {2.69 \, {\rm
GHz}}})^{-0.73 \pm 0.08} ,\eqno\autnum $$

\noindent where we have used the geometric mean between the high frequencies
as reference.  The deviation of the observed flux densities from this fitted
powerlaw is 2 to 3 $\sigma$ for each individual frequency only, but at all
four lower frequencies this deviation is in the same direction, making it of
interest to pursue; to have such a deviation occurring systematically is
significant.  Since systematic errors enter the individual errors, it is
difficult to measure the significance of this systematic difference, but
assuming the systematic errors to be fully accounted for in each individual
error determination, then the combined significance is high, much higher than
3 $\sigma$. In the following this observation is the key to the arguments we
will be presenting.

\titleb {Interpretation}

What is the physical origin for this downturn at low frequency?  It is very
difficult and quite unconvincing to attribute the cutoff to free-free
absorption,  or synchrotron self-absorption (Seaquist et al. 1989).  We
note that Seaquist et al. (1989) already mentioned the possibility of a low
energy cutoff in  the energetic  electron distribution.  A low energy cutoff
in the electron distribution is a natural consequence of nucleus-nucleus
interaction betwen cosmic ray particles and interstellar matter, primarily
proton-proton (pp) interaction; this  is due to the large rest mass of
charged pions which are created in such collisons (neutral pions decay into
photons, of course), and which then decay further,  ultimately leaving
neutrinos of various flavors and electrons and positrons.  This topic of the
injection of secondaries from energetic particle interaction with matter has
been of classical interest (Hayakawa  \& Okuda  1962; Ginzburg \& Syrovatskii
1964; Ekspong et al. 1966; Stecker 1971; and  many others; recent references
are Klemens 1987, Biermann \& Strittmatter  1987, Sikora et al. 1987 and
Mannheim 1993a, b).

The kinematics of the creation of a new particle like one or several charged
pions in the center of mass frame of the encounter of the two protons reduce
the energy available for the pion when the excess energy comes down to the
scale of the proton rest mass, and so we expect the electron/positron
distribution to deviate slightly from the ultimate power law below energies
of a few GeV.  Such a change is indeed observed in proper Monte-Carlo
calculations  (Daum \& Biermann, in prep.).  The peak of the production
spectrum, when multiplied by energy to find the  relevant maximum for
synchrotron emission,  is near 100 MeV:  $\gamma_{e,min} \approx 200.$
Thus the final electron/positron spectrum is the well known power law above a
particle energy of order 5 GeV, but shows a low energy turnover and cutoff at
about 100 MeV. Between 100 MeV and about 5 GeV the spectrum is  somewhat
flatter than the ultimate high energy spectrum.

Hence, the expected radio spectrum is a power law $S_{\nu} \sim \nu^{+1/3} $
in frequency $\nu$ below

$$\nu_{max} \, = \, 7.2 \, 10^{8} \, {\rm Hz} \, ({\gamma_{e,min} \over 200})^2
\,  \sin \theta_{cone} \, B_{0.5},\eqno\autnum $$

\noindent where we have referred the magnetic field at the actual location of
the shock to the magnetic field at the equator at the reference radius by
$B(shock)  =  1.1 \, 10^{-3} \, \sin \theta_{cone} \,B_{0.5}$ from the
assumption of a Parker wind topology; $B_{0.5}$ is the strength of the magnetic
field at the equator at the reference radius of $10^{14}$ cm in units of 3
Gauss. $\theta_{cone}$ is the cone opening angle of the emission region.

The low frequency turnover is in the range $0.3 - 1.4$ GHz, and so this
suggests a magnetic field strength of about 1.3 to 5.8 Gauss at the reference
radius, and more if the cone angle of the observed emission is small.  Here the
latitude dependence of the magnetic field strength enters only linearly, and so
the magnetic field implied here is the lower limit for the equatorial magnetic
field strength.   We thus estimate from the upper as well as lower limit the
magnetic field to  be of order 1 Gauss at the reference radius.

Above the frequency for the maximum in emission (corresponding to
$\gamma_{e,min}$) the radio spectrum has the approximate form $S_{\nu} \sim
\nu^{-0.7},$  for $U_1/V_W \gg 1$, but somewhat steeper than this for smaller
shock  velocities, with some steepening for the frequency  range a factor of
order  3000 above the frequency of maximum emission.  Here $U_1$ is the shock
velocity in the wind frame, and $W$ is the wind velocity.  We emphasize that a)
this spectrum is optically thin, and b) the spectrum is very broad at the
maximum of the flux.

We have used here the Fermi model for the multiplicity of the secondary pions;
there is a weak modification for the spectrum of the secondaries when one
considers the full kinematics of the encounter of the pp-collison, measured
multiplicities and the thresholds (Daum \& Biermann, in prep.).

This spectrum is valid for primary proton spectra of $E^{-7/3}$ which have been
derived for strong shocks which are much faster than the wind; for steeper
proton spectra such as expected for shocks which are slow compared to the
wind the resulting secondary spectrum is also steeper, and so we can reconcile
the observed radio spectrum and its error range with this suggestion.

This spectral behaviour of two well-specified power laws around a peak is thus
{\it a characteristic signature of hadronic interactions}, and provides a
clear statement:  {\it Acceleration of nuclei, including protons, is clearly
indicated by such a spectrum}.  Then there ought to be corresponding
$\gamma$-ray emission, as well as neutrino emission, which, however,
usually is too weak to be detected.  We note that the threshold for new
particle production is a common signature for hadronic interactions, see, e.g.,
also Mannheim (1993a) leading quite generally to breaks between two different
power laws.  Whether such processes also help understand $\gamma$-ray bursts
(e.g., Band et al. (1993)), which also seem to show such {\it nonthermal}
spectra, is open at present.

\titlea {Injection of secondary particles}

\titleb {Basic balance of injection and loss}

In this main section of the paper we will derive the equivalent
injection efficiency of secondary particles from p-p interactions in order to
see whether we can explain the radio data and their implications for i) OB
and WR stars, ii) radio supernovae, and most importantly, iii) GK Per.  The
basic question is to ask is whether we can explain the rather
different injection efficiencies as due to primary injection from the thermal
pool in some cases, and as secondary injection in all other cases; the test is
the quantitative estimate of the efficiency derived and, also, obviously the
spectrum of the particles.

We use the results and notation from CR II and define

$$\eta_p \, \rho \, U_1^2 \; = \; \int_{t_a}^{t_b} C_p \, t^{-7/3} \, E_{kin}
\, d t .\eqno\autnum $$

\noindent with $t := p_p/(m_p c)$; here $p_p$ is the momentum of a
proton, $E_{kin}$ its kinetic energy, $U_1$ the shock speed, and
$\eta_p$ an efficiency of converting shock kinetic energy to an energetic
particle population.  The integration limits $t_a$ and $t_b$ are
from the beginning of the energetic particle population close to the
thermal Maxwellian tail, to the maximum achievable;  as long as $t_a \ll 1$
and $t_b \gg 1$ the integral is independent of their exact values.  Below we
will refer to the equivalent efficiency for electrons $\eta_e$ directly
obtained from observations, and our model for the emission region (see paper
CR II).  The density $\rho$ above is the density in the wind of speed $V_W$
and so we have

$$C_p \; = \; 26.9 \, \eta_{p,-1} \, ({\dot{M}_{-5} \over V_{W,-2}}) \,
U_{1,-2}^2 \, {1 \over r_{14}^2} .\eqno\autnum $$

\noindent The units for $C_p$ are $cm^{-3}$, $10^{-5} \, M_{\odot} /{\rm yr}$
for the mass loss, $10^{14}$ cm for the radius, 0.1 for $\eta_p$, and
$10^{-2} \, c$ for the velocities.  Then interactions  between energetic
protons and thermal protons  in the wind produce secondary electrons and
positrons as a rate per unit  volume

$$C_p \, \int \gamma_p^{-7/3} \, \langle{\sigma_p \, f(\gamma_p)}\rangle \,
c \, n \,  d \gamma_p ,\eqno\autnum $$

\noindent where $f(\gamma_p)$ is the multiplicity.  The combination of
cross section and multiplicity $\langle{\sigma_p \, f(\gamma_p)}\rangle$ we
approximate as  a constant of value $40 \, 10^{-27}\, f_{\Delta}$ cm$^2$;
there are various calculations of this in the classical literature, e.g.
Gould \& Burbidge (1968), Stecker (1971), Stephens \& Badhwar (1981); we
take the particular value from a calculation by Niemeyer (1995), which
improves upon Gould \& Burbidge (1968) and agrees with Stecker (1971), and
obtain an estimate for $f_{\Delta}$ of $1.4$.  As a lower threshold we use
the approximation $\gamma_p = 1.4$ from the threshold condition for the
Delta-resonance (derived from the energy where  the cross section falls to
half its peak value).

The rate of generation of secondary electrons/positrons is then

$$1.53 \, 10^{-7} \, \eta_{p,-1} \, f_{\Delta} \, ({\dot{M}_{-5} \over
V_{W,-2}})^2 \, U_{1,-2}^2 \, {1 \over r_{14}^4} .\eqno\autnum $$

\noindent Here we have made no distinction between hydrogen and other nuclei
in the gas of the stellar wind, but have just counted all nucleons.

We consider the shock frame as it advances.  Then we also have losses, first
of all dilution losses - from the continuity equation in spherical
coordinates - with the rate

$$ 2 \, {{V_W + U_1} \over r} ,\eqno\autnum $$

\noindent and the losses downstream from the diffusive Fermi process (see
Drury 1983) of

$$(1 - \zeta) \,{U_1 \over r} ,\eqno\autnum $$

\noindent in the limit, used here throughout, that the shock velocity ratio
$U_1/U_2 = 4$; this is implied by our assumption that the shock is highly
supersonic. The term $\zeta$ describes the resupply of secondary particles
from other sources further upstream; $\zeta = 1$ denotes full resupply,
i.e., no net loss, and $\zeta = 0$ is for no resupply.

These processes apply to the number of energetic secondaries, which is given
by

$$C_{ep} \int \gamma_{ep}^{-22/9} \, d \gamma_{ep} ,\eqno\autnum $$

\noindent where the integration limits are $\gamma_{ep 1} \, \approx \, 200$
and $\gamma_{ep 2}  \gg  \gamma_{ep 1}$.  We concentrate on the number  of
secondary particles, since further acceleration can not change this number,
but modifies the spectrum.  This can thus be written as

$$F \, = \, {9 \over 13} \, C_{ep} \, \gamma_{ep 1}^{-13/9} .\eqno\autnum $$

Replacing the radius $r$ with the corresponding time dependence we have

$$r_{14} \, = \, 6 \, 10^{-6} \, (U_{1,-2} + V_{W,-2}) \, t ,\eqno\autnum $$

\noindent we have with

$$\eqalign{A \, =& \, 1.18 \, 10^{14} \, \eta_{p,-1} \, f_{\Delta} \cr &
({\dot{M}_{-5} \over V_{W,-2}})^2  \, {1 \over U_{1,-2}^2} \, {1 \over {(1 +
{V_W/U_1})^4}} ,} \eqno\autnum $$

\noindent the balance equation

$${{d \, F} \over {dt}} \, = \, {A \over t^4} \, - \, \epsilon \, {F \over t}
,\eqno\autnum $$

\noindent where

$$\epsilon \, = \, 2 + {{1 - \zeta} \over {1 + {V_W/U_1}}} .\eqno\autnum $$

Obviously, we have $2  \leq  \epsilon \, \leq 3$; the limit of $U_1/V_W
\gg 1$ and $\zeta = 0$ corresponds to $\epsilon = 3$.  The integration yields

$$F \, = \, {1 \over {3 - \epsilon}} \, {A \over t_o^3} \,
(({t_o \over t})^{\epsilon} \, - \, ({t_o \over t})^3) ,\eqno\autnum $$

\noindent for $\epsilon \not= 3$ and

$$F \, = \, {A \over t^3} \, \ln{t \over t_o} ,\eqno\autnum $$

\noindent for $\epsilon = 3$.

Now translating this expression back into a radial dependence we obtain

$$\eqalign{F \, =& \, 2.55 \, 10^{-2} \, {1 \over r_{14,o}^3} \,
(({r_o \over r})^{\epsilon} \, - \, ({r_o \over r})^3)  \,
{1 \over {3 - \epsilon}} \cr & f_{\Delta} \,
\eta_{p,-1} \, ({\dot{M}_{-5} \over V_{W,-2}})^2  \, U_{1,-2} \,
{1 \over {1 + {V_W/U_1}}} ,} \eqno\autnum $$

\noindent and a corresponding expression in the two cases of $\epsilon$
$\not= 3$ and $= 3$.

 From this expression we can derive an equivalent value of $\eta_{ep}^{\star}$
which represents the energy of this population which we derive from
observations ignoring the lower energy cutoff.  The true energy content is
lower by about a factor of 15 because of the low energy cutoff (for
$\gamma_{ep,eff} = 200$).  Using $\gamma_{ep,eff} = \gamma_{ep,1} = 200$ and
$f_{\Delta} = 1.4$ this  corresponds to an equivalent efficiency for
electron/positron injection of

$$\eqalign{\eta_{ep}^{\star} \, =& \, 3.5 \, 10^{-4} \,  {1 \over r_{14,o}} \,
(({r_o \over r})^{\epsilon - 2} \, - \, {r_o \over r})
\, {1 \over {3 - \epsilon}} \cr &
\eta_{p,-1} \, ({\gamma_{ep,eff} \over 200})^{13/9} \,
({\dot{M}_{-5} \over V_{W,-2}})  \, {1 \over U_{1,-2}} \,
{1 \over {1 + {V_W/U_1}}} ,} \eqno\autnum $$

\noindent and again a corresponding expression in the two cases for the value
of $\epsilon$. This is, of course, only a good approximation as long as
$\gamma_{ep,eff} \gg 1$.

The newly injected electrons and positrons are also subject to local energy
gain and losses, which flatten their spectra to an index of $-7/3$ with the
same low energy cutoff (see Drury 1983, section 2.3.2, with an added
integration over the various source energies).  As a consequence the numerical
factors above change only slightly.

In the case that particle interaction is strong initially, say when the
shock is close to the material of the accretion disk, then the formulae
above change in an obvious way.  In the balance equation the term
corresponding to the newly created particles is zero and only an initial
population exists which is slowly being diminished. There is an initial
equivalent efficiency  and thereafter it changes radially as

$$\eta^{\star}_{ep} \; = \; \eta^{\star}_{ep,o} \,
({\gamma_{ep,eff}(r) \over 200})^{13/9} \,
({r_o \over r})^{\epsilon - 2} .\eqno\autnum $$

In the case that there is a continuous or quasi-periodic resupply of energy
to the population of secondary particles $\zeta \approx 1$, this means simply
that $\eta^{\star}_{ep}  =  \eta^{\star}_{ep,o} $ where
$\eta^{\star}_{ep,o}$ can approach the energy
density of primary particles $\eta_{ep,o}  \approx  \eta_p$.  A limit is
$\gamma_{ep,eff} (r)  =  const  \approx  \gamma_{ep,1}  .$

The induced numerical value for $\eta_{ep}^{\star}$ has to be compared with
data.  Therefore we have to ask three  questions: First, can we explain the
estimated value of $\eta_e$ of order $10^{-6}$ for low velocity shocks in
OB and WR  stars, can we interpret the value of about 0.1 in SN-wind shocks,
and can we at the same time interpret  the value required by the observations
of GK Per, of order nearly unity?  This will then also give information on
what to predict in secondary/primary ratio  for high energy electrons.
In order to do this we need to consider first the injection of protons.

\titleb {Injection of energetic protons}

\titlec {The perpendicular case}

Therefore we have to discuss the injection for protons
(and nuclei), since we need to derive a physical theory for the value of
$r_o$, where the protons start getting injected efficiently into the
acceleration process (see Bell 1978a, b, Drury 1983).  Consider a proton
in the post-shock thermal tail of the Maxwellian distribution.  In many papers
Ellison and collaborators have shown that injection from the thermal tail
downstream is possible (see, e.g., Ellison et al. 1990, Jones  \&
Ellison 1991).  We assume that a substantial population of such particles is
built up.

We first consider the case where the unperturbed magnetic field is nearly
parallel to the shock front; this case is valid over most of $4 \pi$ for
a spherical shock going through a Parker wind in its limiting configuration.
In a Parker wind there is a polar cone region, where the magnetic field
configuration is radial, and thus is perpendicular to the shock front.  We
consider this second case in the following subsection.

Now we ask the question:  Can the injection
of particles be maintained to  keep on producing the energetic particle
population?  Consider for the sake of an example the case of a proton:  Such
a proton is subject most of all to two-body encounters
with the thermal electrons, also simply referred to as ionization losses.  The
slow and steady energy  gain from Fermi acceleration  including drifts is
hindered when the energy loss during one acceleration  cycle is equal to or
larger than the energy gain per cycle.  The rate of energy loss is given in,
e.g., Nath \& Biermann (1993), and this rate has to be multiplied by the
residence time in order to obtain the loss per cycle. In  the subrelativistic
regime the change in velocity $\beta c$ can be written as

$$(\Delta \beta)_{loss} \; = \; 1.02 \, 10^{-3} \, ({\dot{M}_{-5} \over
V_{W,-2}})  \, {1 \over U_{1,-2}^3} \, r_{14}^{-1} ,\eqno\autnum $$

\noindent where we have used a factor of $1.3621 \, m_p$ between mass density
and  hydrogen density and a factor of 1.181 between hydrogen density and
electron density, valid for a fully ionized gas at cosmic abundances
(Schmutzler 1987).

The gain per cycle in Fermi acceleration can be written in the subrelativistic
limit as $(\Delta \beta)_{gain}  =  10^{-2} \, U_{1,-2} .$
Putting now gain and loss equal gives a lower limit for the radius,
where  shock acceleration can be sustained, $r_o$:

$$r_{o,14} \; = \; 0.10 \, ({\dot{M}_{-5} \over V_{W,-2}}) \,
{1 \over U_{1,-2}^4} .\eqno\autnum $$

\noindent Putting this back into the expression for the equivalent efficiency
$\eta_{ep}^{\star}$ we obtain for $\epsilon  \not=  3$

$$\eqalign{\eta_{ep}^{\star} \, =& \, 4.8 \, 10^{-4} \,
(({r_o \over r})^{\epsilon - 2} \, - \, {r_o \over r}) \cr &
({\gamma_{ep,eff} \over 200})^{13/9} \,
\eta_{p,-1}  \, U_{1,-2}^3 \, {1 \over {1 + {V_W/U_1}}} ,} \eqno\autnum $$

\noindent and correspondingly for $\epsilon  =  3$, where we have also
included now the small correction for secondary  acceleration of the secondary
particles.

\titlec {The parallel case}

A contrast to the picture above is injection in the polar cone region, because
there the magnetic field is predominantly radial already in its unperturbed
stage. In the injection argument this means that we have to replace the
residence time obtained from the concept of fast turbulent convection with a
residence time derived from a turbulent wave field of waves parallel to the
magnetic field.

The residence time (see, e.g., Drury 1983) is given by

$$\tau_{res} \; = \; {{4 \kappa} \over {U_1 \beta c}} ,\eqno\autnum $$

\noindent where $\kappa$ is the scattering coefficient, $\beta$ is the particle
velocity in units of the speed of light $c$, and $U_1$ is again the shock
velocity.  In the limit of strong turbulence at all relevant wavenumbers the
scattering coefficient is given by

$$\kappa \; = \; {1 \over 3} \, {{\beta m_p c^2} \over {e B(r)}} \, \beta c
,\eqno\autnum $$

\noindent where $B(r)$ is the radial magnetic field, dominant in the polar
cone region region.  We use here the nonrelativistic limit.  From Parker's wind
theory  we have for the polar cone region, that

$$B(r) \; = \; 3 \, 10^{-3} \, {\rm Gauss} \, B_{0.5} \,
{{V_W} \over {\Omega  R_s}} \,  {R_{s,11} \over { r_{14}^2}} ,\eqno\autnum $$

\noindent where $\Omega$ is the angular rotation rate of the star, $V_W$
is again the wind velocity, and $R_{s,11}$ is the radius of the star in
units of $10^{11}$ cm.  The magnetic field topology changes from a
predominantly radial configuration near the equator to an increasingly
tangential configuration at the radius  given by $V_W / \Omega$.  As a
consequence the residence time can be written in this limit as

$$\tau_{res} \; = \; 4.6 \, {\rm sec}\, {\beta \over {B_{0.5} U_{1,-2}}}
\, {{\Omega R_s} \over V_W} \, {1 \over R_{s,11}} \, r_{14}^2 .\eqno\autnum $$

This time is so short because we use the extreme limit of the Larmor radius of
the particles for the characteristic length of the transport.

The loss against interaction with the post-shock hot electrons is the most
effective interaction, and peaks at that velocity for the particles when the
protons have the same velocity as the heated thermal electrons.  As before
we use post-shock parameters for the loss arguments.  Thus the interaction is
maximized at $\beta  =  0.23 \, U_{1,-2} .$  The interaction leads then
to a loss of momentum $\Delta \beta$ equal to the gain from the Lorentz
transformation of a particle going through the shock and getting scattered
back at the condition

$$2.44 \, 10^{-6} \, {{\dot{M}_{-5}} \over V_{W,-2}} \, {1 \over U_{1,-2}^3}
\, {{\Omega R_s} \over V_W} \, {1 \over R_{s,11}} \, = \, 1 ,\eqno\autnum $$

\noindent which is {\it independent} of radius; this is because the quadratic
dependence of density and magnetic field in the polar cone region cancel here.
If this condition is met, then the injection of energetic protons is hindered,
and secondaries cannot be produced.

\titleb {Application to stars}

\titlec {OB and WR stars}

First, let us consider the normal OB and WR stars, where we have argued
(CR II, and Biermann 1995) that $\eta_e$ is of order $10^{-6 \pm 1}$.

It is clear that for ordinary parameters for OB and WR stars the condition
derived above, eq. (27), is never met.  One can show analoguously that
ionization  losses and pp-interaction cannot readily limit particle
acceleration in the polar cone region.

This entails that injection of secondaries from pp-collisions may start quite
close to the star, where densities are high.  The radial behaviour of
$\eta_{ep}^{\star}$ is estimated from our expressions earlier with $\epsilon
\approx 0.5$:

$$\eqalign{\eta_{ep}^{\star} \, =& \, 5.5 \, 10^{-3} \,  {1 \over R_{s,11}} \,
({R_{s,11} \over r_{14}})^{1/2} \cr &
\eta_{p,-1} \, ({\gamma_{ep,eff} \over 200})^{13/9} \,
({\dot{M}_{-5} \over V_{W,-2}})  \, {1 \over U_{1,-2}} ,} \eqno\autnum $$

\noindent suggesting that secondaries {\it do not} play a dominant role
in the radio emission of OB and WR stars; the number is too small because it
refers {\it only} to the polar cone region.  Inverse Compton losses further
diminish any role of these secondaries.

The critical Alfv\'enic Mach number for shocks to inject primary
electrons into an acceleration process in a shock is in the range of 3 to 30 -
the value is highly uncertain at present (see paper CR III).  We estimate the
equatorial Alfv\'enic  Mach number of shocks in the winds of OB and WR stars
to be of order 0.3.   Hence the angular range of electron injection is limited
by $\theta_{cone}   \approx   0.01 \, {\rm to} \, 0.1 $.  Using then the
implied  conal area which participates in the radio emission, and the latitude
dependence of the magnetic field of the Parker model wind, we arrive at the
overall efficiency, i.e. that efficiency which we infer from
observations ignoring the angular limitation.  This number is then in the
range $10^{-6}$  to $10^{-8}$ as required by the observations.

Our conclusion here then is that primary injection in a polar cone can
explain the low apparent efficiency  of the nonthermal radio emission
in OB and WR stars.  We emphasize that we consider single stars here.  Given
sufficient flux density, this conal restriction may be observable with VLBI.

\titlec {Radio supernovae}

Second, we can ask what is expected for radio supernovae:  Here we likely
have  $\epsilon = 3$, $U_{1,-2} = 3$.  The observed radio supernovae all
except 1987A seem to have exploded into red giant winds with low velocity with
$V_{W,-2} \approx 0.01$. It is easy to see that the radio emission from
radio supernovae is fully consistent with all energetic electrons being
primarily accelerated electrons which dominate by about two orders of
magnitude  or even more over the secondary electrons/positrons.  However, when
the  supernova shock hits the surrounding shell that the steady wind often
builds up  over the lifetime of the star, then substantial secondary particle
production takes place, which dominates over the earlier history
of the shock running through the wind (see, e.g., Nath \& Biermann 1994).
We will discuss cosmic ray electrons elsewhere (paper CR VI, in prep.).

\titlec {GK Per}

Last, let us consider the case of the nova GK Per: Optical data on the
emitting mass, usually identified with the  ejected mass, (Pottasch 1959)
give a mass of $7.0 \, 10^{-5} \, \rm M_{\odot}$.  We identify this
with the mass in the shocked shell of the wind material, under the
premise that the ejecta caused a shock to run through a  pre-existing wind.
With this notion we obtain for   $ \dot{M}_{-5} / V_{W,-2}$ a value of
$0.22 \, f_{\rho}$.  However, nova  shells are not spherically symmetric and
so the spherically averaged mass  in the wind  shell is estimated to be
approximately three times higher  (Seaquist et al.  1989).  We will use
this higher value in the following,  and use $f_{\rho}$ to account for any
further differences.  Using such an approach the uncertainty in $r_o$
becomes so large, that we keep this radius  in the expression  and  so to
a fair approximation we have for GK Per

$$\eqalign{\eta_{ep}^{\star} \, =& \, 2.3 \, 10^{-3} \,  {1 \over
r_{o,14}^{3-\epsilon}} \,  {1 \over r_{14}^{\epsilon-2}}
\cr &  ({\gamma_{ep,eff} \over 200})^{13/9} \,  \eta_{p,-1} \, f_{\rho} \,
{1 \over U_{1,-3}}.} \eqno\autnum $$

The lowest reasonable limit for the injection radius is clearly the size of
the binary system, or the inner region of the accretion disk, giving $r_{o,14}
 > 4 \, 10^{-3}$ and $r_{o,14}  >  4 \, 10^{-4}$.  With the full
limitation of the observed radio emission to a cone  and no allowance for
clumping in the magnetic field we find that the efficiency can indeed
be as high as required by the observations, under the condition that the shock
speed is smaller than the Alfv\'en velocity at the equator.

If we try to find  solutions where there is appreciable secondary production
in the wind, we fail; the main injection, in this picture, has to arise from
the inital environment, at the inner edge of the accretion disk.

The source parameters derived here suggest that
secondaries from hadronic interactions, i.e. p-p interactions here, give an
equivalent injection efficiency between $\approx \, 10^{-2}$ and 1.  All this
implies, that $\gamma_{ep,eff}$ is approximately constant, which leads to a
solution close to the case $\eta_{ep}^{\star} \, \approx \, const$ and
maximal, and $\epsilon$ very close to a value of 2.  This in turn implies
either a slow shock ($V_W \gg U_1$) or substantial resupply of secondaries
from upstream ($\zeta \approx 1$).

Hence we have found a consistent, albeit speculative picture that explains
the radio emission of GK Per as produced by secondary electrons/positrons,
with a theory which  also gives acceptable estimates for OB and WR stars,
as well as radio supernovae, where the {\it standard} acceleration from
the thermal pool is probably dominant, but restricted to the polar cone as an
injector for OB and WR stars.

\titlea {Consequences for other sources and other wavelengths}

\titleb {Other stellar binary systems}

The model implies copious pp-interaction at the beginning of the shock
travelling through the wind.  A consequence is both a power law continuum from
$\pi^o$ decay as well as an annihilation line from secondaries.  In nova
Muscae Chen et al. (1993) have observed a 511 keV annihilation line, which
they argue arises from a region near the inner edge of an accretion disk.
In the same nova Goldwurm et al. (1992) describe observations of a
$\gamma$-continuum with slope $-2.38 \pm 0.05$, consistent with a primary
spectrum of protons of slope near $-7/3$.

\titleb {Other radio sources with related characteristics}

There are a number of other radio sources which appear to show properties
consistent with the picture which we propose.  Secondaries have been discussed
to explain other features in the spectra of radio sources (see, e.g.,
Klemens 1987, Biermann \& Strittmatter 1987, Mannheim 1993a, 1993b).

Klemens (1987) described the X-ray emission of the M87 jet
successfully with the concept that secondaries from hadronic interaction give
rise to synchrotron X-ray emission.

The Galactic center source shows a spectrum which rises approximately as
$\nu^{1/3}$ through the radio range; it is not clear at this stage whether
this spectral behaviour is due to optical thickness effects or due to another
process such as discussed here.  The Galactic center source has been discussed
elsewhere (Falcke 1995), following our earlier modelling (Falcke et al. 1993,
Falcke \& Biermann 1995, Falcke et al. 1995).

The most prominent example among radio galaxies is Cyg A (Muxlow et al. 1988).
The radio emission of the lobes shows a low frequency turnover near 200 MHz
which then can be connected with an estimate of the relevant magnetic field
to a minimum energy of the electron distribution of about
$\gamma_1 \simeq 350 \pm 50$.  There is evidence that radio galaxies in general
tend to flatten towards low frequency (Laing \& Peacock 1980,
Mangalam \& Gopal-Krishna 1994), which may be related to a low energy
cutoff in the electron distribution; the effect should be strong only at very
low radio frequencies due to the estimated low magnetic field strengths in
radio galaxies.

\titlea {Is there an alternative?}

Where is the main weakness of the model proposed?  The key is clearly
the argument that hadronic secondaries are able to account for the spectrum.
Therefore, in this section we propose to ask whether primary injection and
acceleration can also produce the pattern of radio polarization, radio
emission strength, radio structure and radio spectrum.

The radio polarization has been argued to arise from the instability of a
cosmic ray mediated shock (see paper CR I, CR II and Biermann 1995).  The
argument  on secondary or primary injection is indifferent to this, since
we use primary protons to drive the instability.

The radio emission strength could be explained by invoking little or no
magnetic clumping and just using the decrease of total flux from the
geometry of the underlying magnetic field.  Since the magnetic field
strength in a Parker wind decreases  as $\sin \theta$ towards the poles,
the total radio emission is reduced by a factor of one to two orders of
magnitude, depending upon which cutoff angle we use.  This entails a
corresponding increase in the implied efficiency to about the maximum
possible.  The cutoff angle would still arise from the limiting condition on
injection, and hence correspond to a lower magnetic field strength than used
above.  This runs into the problem that the efficiency may then have to be
increased beyond any reasonable level.

The radio structure would derive from the same argument as before, the drift
asymmetry in the accelerated particle population in the two hemispheres (see
paper CR I, CR II and Biermann 1995).

The radio spectrum would also be easy to accomodate in the concept that
the explosion advances into the interstellar medium (paper CR III); however,
the spatial asymmetry is then difficult to interpret.

Finally, for the low frequency cutoff we would require a new and rather
difficult explanation, like foreground absorption, an extremely clumpy
absorbing medium, or something else.  Seaquist et al. (1989) discussed all
likely possibilities, and found no other reasonable explanation.  Therefore
the radio observation of the low frequency turnover is the key
observation on which everything rests.  A very low frequency spectrum of
GK Per would be decisive to settle the interesting question on the role of
hadronic secondaries, and thus on the presence of accelerated protons in a
specific radio source.

\titlea {Conclusion}

We have tested a model proposed earlier to account for the spectrum, particle
energies and chemical composition of cosmic rays with the observations of the
nonthermal radio emission from the nova GK Per.  We obtain a consistent
picture implying a pre-existing strong stellar wind.  Such a
strong wind can strongly influence the evolution of a low mass binary like GK
Per due to mass and angular momentum loss from the system (see, e.g., Kim et
al. 1992, Iben 1995).

We interpret the radio spectrum of GK Per as resulting from
hadronic secondary electrons/positrons, which, in the optically thin case,
give a characteristic radio spectrum with a spectral index of +1/3 below a
maximum and, for a strong shock in a wind slow compared to the shock speed,
of $\approx  -0.7$ above the maximum.  We argue that such a radio spectrum
is a unique characteristic for hadronic interactions.  We can account for
the observed  radio emission, its spectrum including the low frequency cutoff,
its radio polarization properties, geometric shape and its luminosity.

We have applied the concepts developed to OB and Wolf Rayet
stars, as well as radiosupernovae, and have been able to verify that the
efficiencies of injection implied by the observations can be readily
interpreted quantitatively, however, with primary injection.  One consequence
is that the radio emission in OB and Wolf Rayet stars is from a polar cone,
just as in GK Per, but restricted much more in angle.

We have shown, that the radio emission properties of the old nova GK Per can
consistently be interpreted in the context of our proposed model for the
origin of cosmic rays, and have made a small step in the direction of
accounting for secondaries from nucleus - nucleus interaction in the
populations of energetic particles.  We point out the key radio observation to
confirm the model proposed, an accurate very low frequency spectrum of GK Per.

\ack{The authors appreciate the help of Dr. W. Duschl in elucidating the role
and physics of cataclysmic variables and novae; he was an important partner in
starting this paper.  PLB wishes to acknowledge helpful discussions with Drs.
K. Daum, Gopal Krishna, C. Jarlskog, H. Lesch, G. Mann, B.B. Nath, M. Niemeyer,
S.P. Reynolds and E.R. Seaquist.  Critical comments by the referee helped to
bring this paper into focus.  H. Falcke was supported by the DFG under grant Bi
191/9.}

\begref

\ref Abbott, D.C., Bieging, J.H., Churchwell, E., Torres, A.V.:  1986
     Astrophys. J. 303, 239
\ref Baars, J.W.M., Genzel, R., Pauliny-Toth, I.I.K., Witzel, A.:  1977 Astron.
     \& Astroph. 61, 99
\ref Band, D. et al. 1993:  Astrophys. J. 413, 281
\ref Bell, A.R.:  1978a Monthly Not. Roy. Astr. Soc. 182, 147
\ref Bell, A.R.:  1978b Monthly Not. Roy. Astr. Soc. 182, 443
\ref Bieging, J.H., Abbott, D.C., Churchwell, E.B.:  1989 Astrophys.
     J. 340, 518
\ref Biermann, P.L., Strittmatter, P.A.:  1987 Astrophys. J. 322, 643
\ref Biermann, P.L.:  1993 Astron. \& Astroph. 271, 649
     (paper CR I)
\ref Biermann, P.L., Cassinelli, J.P.:  1993 Astron. \& Astroph. 277,
     691 (paper CR II)
\ref Biermann, P.L., Strom, R.G.:  1993 Astron. \& Astroph. 275, 659
     (paper CR III)
\ref Biermann, P.L., Gaisser, T.K., Stanev, T.:  1995 Phys. Rev. D (in press)
\ref Biermann, P.L.:  1995 in "Cosmic winds and the Heliosphere", Eds.
     J.R.Jokipii et al., Univ. of Arizona Press, Tucson, AZ, (in press)
\ref Chen, W., Gehrels, N. Cheng, F.H.:  1993 Astrophys. J. Letters 403, L71
\ref Dickel, J.R., Breugel, W.J.M. van, Strom, R.G.:  1991 Astron. J. 101, 2151
\ref Drury, L.O'C.:  1983 Rep.Prog.Phys. 46, 973
\ref Ekspong, A.G., Yamadagni, N.K., Bonnevier, B.:  1966 Phys. Rev. Letters
    16, 664
\ref Ellison, D.C. et al.:  1990 Astrophys. J. 352, 376
\ref Falcke, H., Mannheim, K., Biermann, P.L.: 1993 Astron. \& Astroph. 278, L1
\ref Falcke, H., Biermann, P.L.:  1995 Astron. \& Astroph. 293, 665
\ref Falcke, H., Malkan, M.A., Biermann, P.L.:  1995 Astron. \& Astroph. 298,
375

\ref Falcke, H.:  1995 in IAU Sympos. 169 ``Unsolved problems of the
     Galaxy", Ed. L. Blitz, Kluwer, Dordrecht (in press)
\ref Ginzburg, V.L., Syrovatskii, S.I.: 1964 ``The origin of cosmic rays",
     Pergamon Press, Oxford
\ref Goldwurm, A. et al.:  1992 Astrophys. J. Letters 389, L79
\ref Gould, R.J., Burbidge, G.R.:  1968 Handbuch der Physik 46/2, p. 265
\ref Hayakawa, S., Okuda, H.: 1962 Progr. Theoret. Phys. 28, 517
\ref Hjellming, R.M.:  1990 in ``Physics of classical novae", Eds. A.
     Cassatella, R. Viotti, Lecture Notes in Physics No. 369, Springer,
     Berlin, p.169
\ref Iben, I. jr.:  1995 Phys. Rep. 250, 1
\ref Jokipii, J.R., Levy, E.H., Hubbard, W.B.:  1977 Astrophys. J. 213, 861
\ref Jones, F.C., Ellison, D.C.:  1991 Space Sc. Rev. 58, 259
\ref Kim, S.-W., Wheeler, J.C., Mineshige, S.:  1992 Astrophys. J. 384, 269
\ref Klemens, Y.:  1987 Ph.D. Thesis University of Bonn
\ref Laing, R.A., Peacock, J.A.:  1980 Monthly Not. Roy. Astr. Soc. 190, 903
\ref MacFarlane, J.J., Cassinelli, J.P.:  1989 Astroph. J. 347, 1090
\ref Mangalam, A.V., Gopal Krishna:  1994 Monthly Not. Roy. Astr. Soc.
     (submitted)
\ref Mannheim, K.:  1993a Astron. \& Astroph. 269, 67
\ref Mannheim, K.:  1993b Phys. Rev. D 48, 2408
\ref Muxlow, T.W.B., Pelletier, G., Roland, J.:  1988 Astron. \& Astroph.
     206, 237
\ref Nath, B.B., Biermann, P.L.:  1993 Monthly Not. Roy. Astr. Soc. 265, 241
\ref Nath, B.B., Biermann, P.L.:  1994 Monthly Not. Roy. Astr. Soc.
     270, L33
\ref Niemeyer, M.: 1995 Ph.D. Thesis University of Bonn
\ref Pottasch, S.R.:  1959 Ann. d'Astroph. 22, 412
\ref Rachen, J.P., Biermann, P.L.:  1993 Astron. \& Astroph. 272, 161
     (paper UHE CR I)
\ref Rachen, J.P., Stanev, T., Biermann, P.L.:  1993 Astron. \& Astroph. 273,
     377 (paper UHE CR II)
%
\ref Reynolds, S.P., Chevalier, R.A.:  1984 Astrophys. J. Letters 281, L33
\ref Romanova, M.M., Lovelace, R.V.E., Bisnovatyi-Kogan, G.S.:  1994 Monthly
     Not. Roy. Astr. Soc. (submitted)
\ref Schmutzler, T. 1987 Ph.D. Thesis University of Bonn
\ref Seaquist, E.R., Bode, M.F., Frail, D.A., Roberts, J.A., Evans, A.,
     Albinson, J.S.:  1989 Astrophys. J. 344, 805
\ref Sikora, M., Kirk, J.G., Begelman, M.C., Schneider, P.:  1987
     Astrophys. J. Letters 320, L81
\ref Stanev, T., Biermann, P.L., Gaisser, T.K.:  1993 Astron. \& Astroph. 274,
     902 (paper CR IV)
\ref Stephens, S.A., Bhadwar, G.D.: 1981 Astroph. \& Space Sci. 76, 213
\ref Stecker, F.W.:  1971  ``Cosmic Gamma Rays", NASA-book
\endref

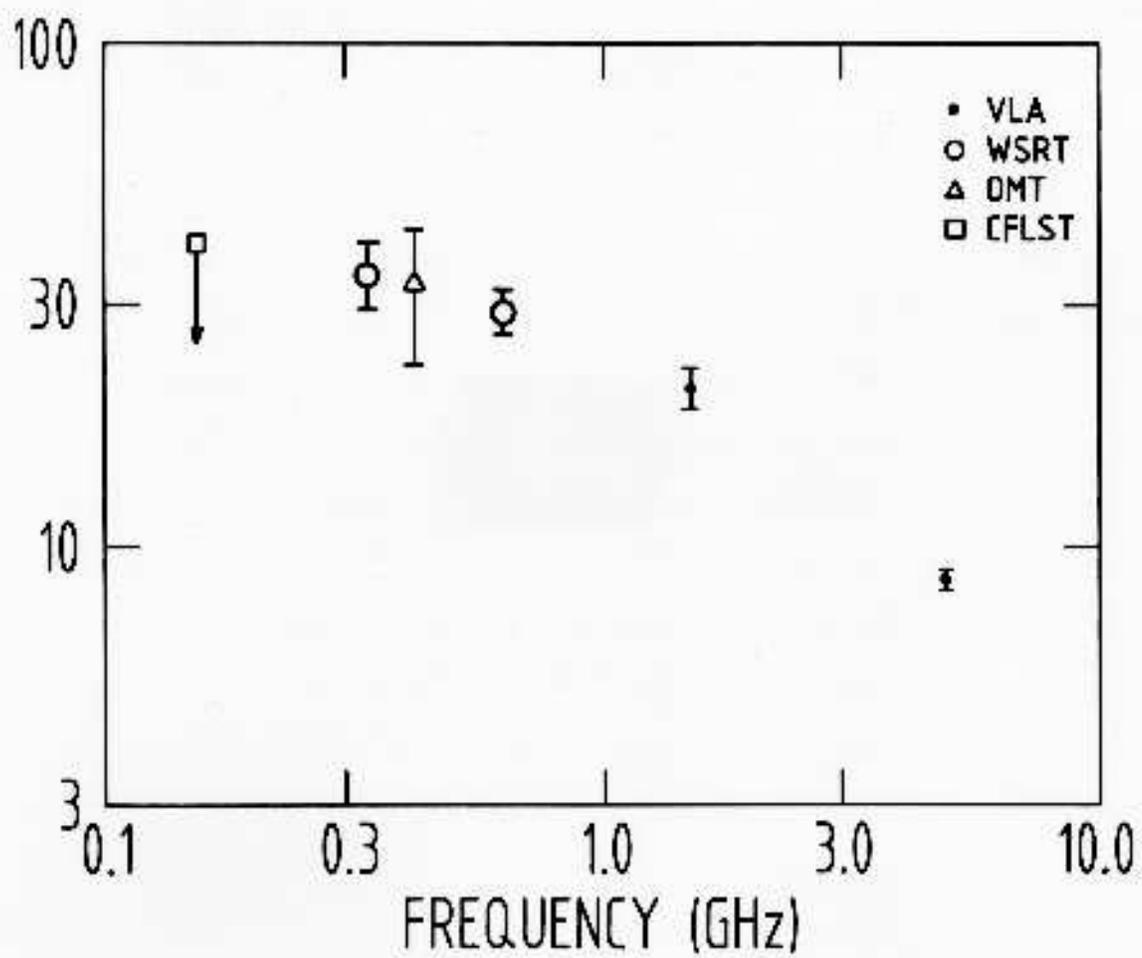
\bye